\documentclass[twocolumn,nofootinbib,amsmath,amssymb,a4paper]{revtex4}

\usepackage{graphicx}
\usepackage{dcolumn}
\usepackage{bm}

\newcommand{\bequ}{\begin{equation}}
\newcommand{\eequ}{\end{equation}}
\newcommand{\bea}{\begin{eqnarray}}
\newcommand{\eea}{\end{eqnarray}}

\newcommand{\bi}{\begin{itemize}}
\newcommand{\ei}{\end{itemize}}
\newcommand{\tev}{\, {\rm TeV}}
\newcommand{\gev}{\, {\rm GeV}}

\def\kpn{K^+\rightarrow\pi^+\nu\bar\nu}

\def\klpn{K_{L}\rightarrow\pi^0\nu\bar\nu}

\begin{document}

\title{Quark and Lepton Flavour Physics\\
in the Littlest Higgs Model with T-Parity}

\author{Andrzej J.~Buras and Cecilia Tarantino}
\affiliation{Technische Universit\"at M\"unchen,
D-85748 Garching, Germany}

\begin{abstract}
The Littlest Higgs model with T-parity (LHT) contains new sources of flavour and CP
violation both in the quark and the lepton sector.
They originate from the
interactions of ordinary fermions with mirror fermions mediated by new gauge
bosons: $W_H^{\pm}$, $Z^0_H$ and $A_H^0$.
The most spectacular departures from the Standard Model are found in $K_L
\to \pi^0 \nu \bar \nu$, $K^+ \to \pi^+ \nu \bar \nu$, in the CP-asymmetry 
$S_{\psi \phi}$ and in lepton flavour violating decays.
In particular, the latter decays offer a clear distinction between the LHT
model and the MSSM.
We summarize the most interesting results of three extensive analyses of
flavour physics in the LHT model.

\end{abstract}

\maketitle

\section{The LHT Model}
One of the most attractive solutions to the so-called {\it little hierarchy
  problem} that affects the Standard Model (SM) is provided by Little Higgs models.
They are perturbatively computable up to $\sim 10 \tev$ and have a rather
  small number of parameters, although their predictivity can be weakened by a
  certain sensitivity to the unknown ultraviolet (UV) completion of the
  theory.
In these models, in contrast to supersymmetry, the problematic quadratic
divergences to the Higgs mass are cancelled by loop contributions of new
particles with the same spin-statistics of the SM ones and with masses around
$1 \tev$.

The basic idea of Little Higgs models\cite{ACG} is that the Higgs is 
naturally light as it is identified with a Nambu-Goldstone 
boson (NGB) of a spontaneously broken global symmetry.
An exact NGB, however, would have only derivative interactions. 
Gauge and Yukawa interactions of the Higgs have to be incorporated. This can
be done without generating
quadratically divergent one-loop contributions to the Higgs mass, through the
so-called {\it collective symmetry breaking}. 
Collective symmetry breaking (SB) has the peculiarity of generating
the Higgs mass only when two or more couplings in the Lagrangian are
non-vanishing, thus avoiding one-loop quadratic divergences.

The most economical, in matter content, Little Higgs model is the Littlest
Higgs (LH) model\cite{ACKN}, where the global group $SU(5)$ is spontaneously broken
into $SO(5)$ at the scale $f \approx \mathcal{O}(1 \tev)$ and
the electroweak (ew) sector of the SM is embedded in an $SU(5)/SO(5)$ non-linear
sigma model. 
Gauge and Yukawa Higgs interactions are introduced by gauging the subgroup of
$SU(5)$: $[SU(2) \times U(1)]_1 \times [SU(2) \times U(1)]_2$. 
In the LH model, the new particles appearing at the $\tev$ scales are the heavy
gauge bosons ($W^\pm_H, Z_H, A_H$), the heavy top ($T$) and the scalar triplet 
$\Phi$.

In the LH model, significant corrections to ew observables come
from tree-level heavy gauge boson contributions and the triplet vacuum 
expectation value (vev) which breaks the custodial $SU(2)$ symmetry. 
Consequently, ew precision tests are satisfied only for quite large
values of the New Physics (NP) scale $f \ge 2-3 \tev$\cite{HLMW,CHKMT}, unable to solve
the little hierarchy problem.
Motivated by reconciling the LH model with ew precision tests, Cheng and 
Low\cite{CL} proposed to enlarge the symmetry structure of the theory by
introducing a discrete symmetry called T-parity.
T-parity forbids 
the tree-level contributions of  heavy gauge bosons and the
interactions that induced the triplet vev.
The custodial $SU(2)$ symmetry is restored and the compatibility with ew
precision data is obtained already for smaller values of the NP scale, $f \ge
500 \gev$\cite{HMNP}.
Another important consequence is that particle fields are T-even or T-odd
under T-parity. The SM particles and the heavy top
$T_+$ are T-even, while the heavy gauge bosons $W_H^\pm,Z_H,A_H$ and the
scalar triplet $\Phi$ are T-odd.
Additional T-odd particles are required by T-parity: 
the odd heavy top $T_-$ and the so-called mirror fermions, i.e.,
fermions corresponding to the SM ones but with opposite T-parity and $\mathcal{O}(1 \tev)$ masses.
Mirror fermions are characterized by new flavour interactions with SM fermions
and heavy gauge bosons, which involve in the quark sector two new unitary 
mixing
matrices analogous to the CKM matrix~\cite{CKM}.
They are $V_{Hd}$ and
$V_{Hu}$, respectively involved when the SM quark is of down- or up-type,
and satisfying $V_{Hu}^\dagger V_{Hd}=V_{CKM}$\cite{HLP}.
Similarly, two new mixing matrices, $V_{H\ell}$ and $V_{H\nu}$, appear in the
lepton sector, respectively involved when the SM lepton is charged or a
neutrino and related to the PMNS matrix~\cite{pmns} through $V_{H\nu}^\dagger V_{H\ell}=V_{PMNS}^\dagger$. 
Both $V_{Hd}$ and $V_{H\ell}$
  contain $3$ angles, like $V_{CKM}$ and $V_{PMNS}$, but $3$
  (non-Majorana) phases \cite{SHORT}, i.e. 
  two additional phases relative to the SM matrices, that cannot be rotated
  away in this case.

Because of these new mixing matrices, the LHT model does not belong to the
Minimal Flavour Violation (MFV) class of models~\cite{UUT,AMGIISST} and 
significant effects in flavour observables are possible, without adding new 
operators to the  SM ones.
Finally, it is important to recall that Little Higgs models are low
energy non-linear sigma models, whose unknown UV-completion introduces a
theoretical uncertainty, as discussed in detail in~\cite{BPUB,BBPRTUW}.
\section{LHT Flavour Trilogy}
Several studies of flavour physics in the LH model without T-parity have been
performed in the last four years\cite{BPUB,FlavLH}. Without T-parity, mirror 
fermions and new sources of flavour and CP-violation are absent, the LH model
is a MFV model and NP contributions result to be very small. 

More recently, flavour physics analyses have also been performed in the LHT
model, for both quark~\cite{HLP,BBPRTUW,BBPTUW} and lepton sectors~\cite{Indian,BBDPT}.
In this model, new mirror fermion interactions can yield large NP effects,
mainly in $K$ and $B$ rare and CP-violating decays and in lepton flavour violating decays. 

Below, we summarize the main results found in our trilogy on FCNC processes in
the LHT model~\cite{BBPRTUW,BBPTUW,BBDPT}.

\subsection{$\mathbf{\Delta F=2}$ Processes in the Quark Sector~\cite{BBPTUW}}

The short distance structure of $\Delta F =2$ processes in the LHT model 
is fully encoded in three perturbatively calculable functions
\bequ
S_K \equiv |S_K| e^{i 2 \varphi_K}\,,\, S_{B_d} \equiv |S_{B_d}| e^{i 2
  \varphi_{B_d}}\,,\, S_{B_s} \equiv |S_{B_s}| e^{i 2 \varphi_{B_s}}\,,
\label{eq:SKds}
\eequ
relevant for $K^0 - \bar K^0$, $B^0_d - \bar B^0_d$ and $B^0_s - \bar B^0_s$
mixings, respectively.

In the SM they all reduce to a real single function $S_\text{SM} = S_0(x_t)$
that is dominated by box diagrams with top quark exchanges.
In the LHT model, where the new mixing matrix $V_{Hd}$ is present, the inclusion of box diagrams with internal mirror quarks and heavy gauge
bosons ($W_H^\pm$, $Z_H$, $A_H$) makes the one-loop functions
in~(\ref{eq:SKds}) complex quantities.
Moreover, the universality between $K^0$, $B_d$ and $B_s$ systems, valid in
the SM is broken so that the magnitudes $|S_i|$ and
the phases $\varphi_i$ depend on $i=K, B_d, B_s$.
We recall that in constrained MFV models~\cite{UUT,BBGT} the short distance functions are  as in the SM
real and universal, although different from $S_\text{SM}$.

The size of $\varphi_i$ depends on the structure of the matrix $V_{Hd}$, which
is so far only weakly constrained by the existing data on $\Delta M_K$,
$\Delta M_d$, $\Delta M_s$, $\sin 2 \beta$ and $\varepsilon_K$, mainly due to
significant hadronic uncertainties in $\Delta M_{d,s}$ and $\varepsilon_K$.
This allows to obtain interesting departures from the SM, in particular in the
$B^0_s - \bar B^0_s$ system but also in connection with the slight
discrepancy~\cite{UTfit,CKMfit,BBGT}, existing within the SM, between the values of $|V_{ub}|$ obtained 
from tree level decays and the value of $\sin 2 \beta$ measured from the 
CP-asymmetry $S_{\psi K_s}$.

The main messages from~\cite{BBPTUW} are as follows:
\begin{itemize}
\item{The presence of a non-vanishing phase $\varphi_{B_d}$ implies that
    $S_{\psi K_s}=\sin (2 \beta + 2 \varphi_{B_d})$, so that with
    $\varphi_{B_d} \simeq -5^\circ$ the possible discrepancy between $|V_{ub}|$
    and $S_{\psi K_s}$ can be cured.}
\item{The presence of a non-vanishing phase $\varphi_{B_s}$ allows to enhance
    the CP-asymmetry $S_{\psi \phi}$ from the SM prediction $0.04$ to $0.30$ with
    an analogous enhancement of the semileptonic asymmetry $A^s_{SL}$ and a
    smaller but sizable enhancement of $A^d_{SL}$.}
\item{A non-vanishing $\varphi_{B_s}$ allows also to slightly suppress $\Delta
    M_s$ below its SM value, thus further improving the agreement with the CDF measurement~\cite{CDFD0}.}
\end{itemize}

\subsection{$\mathbf{\Delta F=1}$ Processes in the Quark Sector~\cite{BBPRTUW}}
\begin{figure}[!]
\includegraphics[width=0.4\textwidth]{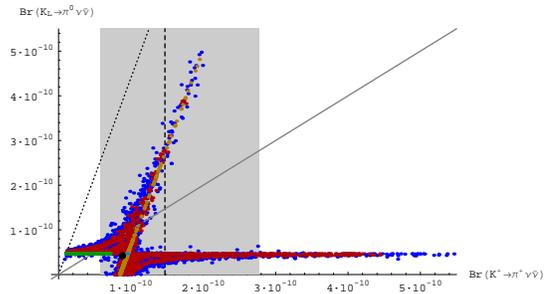}
\vspace*{-0.4cm}
\caption{\label{fig:KLKp} $Br(\klpn)$ vs. $Br(\kpn)$. The shaded
    area represents the experimental $1\sigma$-range for $Br(\kpn)$. The
    model-independent Grossman-Nir bound~\cite{GNbound} is displayed by the dotted line, while the solid line
    separates the two areas where $Br(\klpn)$ is larger or smaller than
    $Br(\kpn)$.}
\label{fig:KLKp}
\end{figure}
The short distance structure of the LHT model, that is relevant for rare $K$
and $B$ decays, is fully encoded in nine perturbatively calculable functions
($i=K,d,s$)
\bequ
X_i = |X_i| e^{i \theta_X^i}\,,\quad Y_i = |Y_i| e^{i \theta_Y^i}\,,\quad Z_i
= |Z_i| e^{i \theta_Z^i}\,,
\label{eq:XYZ}
\eequ
that result from the SM box and penguin diagrams and analogous diagrams with
new particle exchanges.
In the case of the radiative decay $B \to X_s \gamma$, also the function $D'$ has to
be considered.
In the SM and in models with constrained MFV all these functions are real and
independent of $i$, with consequent strong correlations between various
observables in $K$, $B_d$ and $B_s$ systems.  

In the LHT model, the non-vanishing $\theta$'s originate from the new phases 
of the $V_{Hd}$ matrix.
The additional dependence on $i$ in~(\ref{eq:XYZ}) and the possibility of
large $\theta^K_{X,Y,Z}$ imply a pattern of FCNC processes that differs
significantly from the SM and the constrained MFV one.

\begin{figure}[!]
\includegraphics[width=0.4\textwidth]{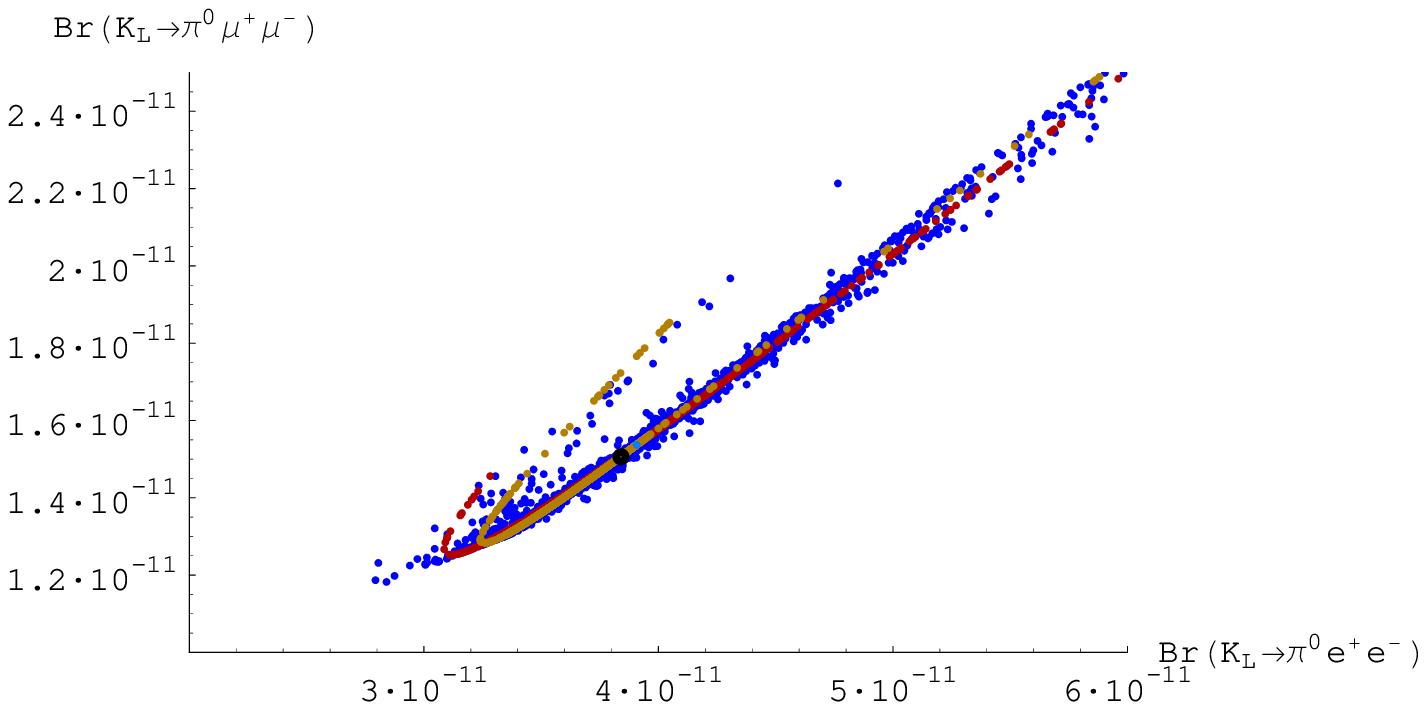}
\vspace*{-0.4cm}
\caption{\label{fig:KmuKe} $Br(K_L \to \pi^0 \mu^+\mu^-)$ vs. 
$Br(K_L\to \pi^0 e^+e^-)$.}
\end{figure}

\begin{figure}
\includegraphics[width=0.4\textwidth]{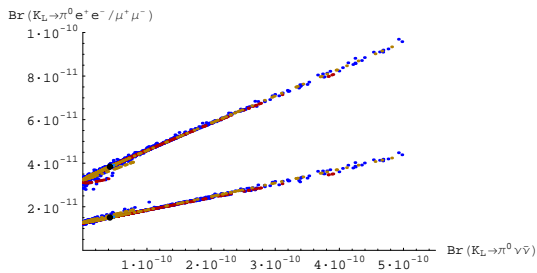}
\vspace*{-0.4cm}
\caption{\label{fig:KLllKLnn} $Br(K_L\to \pi^0 e^+e^-)$ (upper curve) and  $Br(K_L \to \pi^0
  \mu^+\mu^-)$ (lower curve) as functions of $Br(\klpn)$. The corresponding SM
  predictions are represented by dark points.}
\end{figure}

The main messages from~\cite{BBPRTUW} are as follows:
\begin{itemize}
\item{The most evident departures from the SM predictions are found in $K \to
    \pi \nu \bar \nu$ decays (Fig.~\ref{fig:KLKp}).
$Br(K_L \to \pi^0 \nu \bar \nu)$ can be enhanced even by an order of magnitude
    and $Br(K^+ \to \pi^+ \nu \bar \nu)$ by a factor $5$.
Moreover, $Br(K_L \to \pi^0 \nu \bar \nu)$ can be larger than $Br(K^+ \to
    \pi^+ \nu \bar \nu)$, which is not possible in MFV models.}

\item{$Br(K_L \to \pi^0 e^+ e^-)$ and $Br(K_L \to \pi^0 \mu^+ \mu^-)$ can be
    both enhanced by a factor $2-3$ and are strongly correlated, as shown in
    Fig.~\ref{fig:KmuKe}.} 
\item{A strong correlation between $Br(K_L \to \pi^0 \ell^+ \ell^-)$ and
    $Br(K_L \to \pi^0 \nu \bar \nu)$ exists, as shown in Fig.~\ref{fig:KLllKLnn}.} 
\item {The branching ratios for $B_{s,d} \to \mu^+ \mu^-$ and $B \to X_{s,d}
    \nu \bar \nu$, instead,  are modified by at most $50\%$ and $35\%$,
    respectively, and the effects of new electroweak penguins in $B \to \pi K$
    are small, in agreement with the recent data.
The new physics effects in $B\to X_{s,d}\gamma$ and $B\to X_{s,d}\ell^+\ell^-$
turn out to be below $5\%$ and $15\%$, respectively, so that agreement 
with the data can easily be obtained.}
\item{The universality of new physics effects, characteristic for MFV models,
    can be largely broken, in particular between $K$ and $B_{s,d}$ systems.
NP effects, in fact, are typically larger in $K$ system where the SM contribution is
    CKM-suppressed.
In particular, sizable departures from MFV relations between $\Delta M_{s,d}$
    and $Br(B_{s,d} \to \mu^+ \mu^-)$ and between $S_{\psi K_S}$ and the $K \to
    \pi \nu \bar \nu$ decay rates are possible.}
\end{itemize}

\subsection{Lepton Flavour Violation~\cite{BBDPT}}
\begin{figure}
\includegraphics[width=0.4\textwidth]{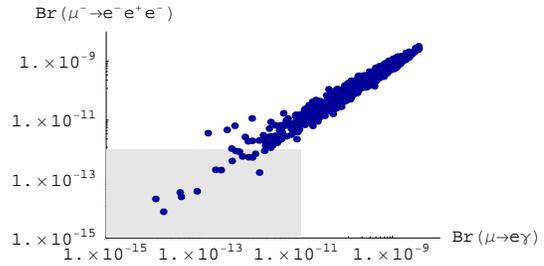}
\vspace*{-0.4cm}
\caption{\label{fig:mue} Correlation between the branching ratios for $\mu\to e\gamma$ and
  $\mu^-\to e^-e^+e^-$. The shaded
  area represents present experimental upper bounds.}
\end{figure}
In contrast to rare K and B decays, where the SM contributions play an
important and often dominant role, in the LHT model the smallness of ordinary
neutrino masses assures that mirror fermion contributions to lepton flavour
violating (LFV) processes are by far the dominant effects.
Moreover, the absence of QCD corrections and hadronic matrix elements allows
in most cases to make predictions entirely within perturbation theory.

In~\cite{BBDPT} we have studied the most interesting LFV processes: $\ell_i
\rightarrow \ell_j \gamma$, $\tau \rightarrow \ell P$ (with $P=\pi, \eta, \eta'$), $\mu^- \rightarrow e^-
e^+ e^-$, the six three-body decays $\tau^- \rightarrow l_i^- l_j^+ l_k^-$ and 
the rate for $\mu-e$ conversion in nuclei.
We have also calculated the rates for $K_{L,S} \rightarrow \mu e$, $K_{L,S}
\rightarrow \pi^0 \mu e$, $B_{d,s} \rightarrow \mu e$, $B_{d,s} \rightarrow
\tau e$ and $B_{d,s} \rightarrow \tau \mu$.

\begin{table}
{\renewcommand{\arraystretch}{1.5}
\begin{center}
\begin{tabular}{|c|c|c|c|}
\hline
ratio & LHT  & MSSM (dipole) & MSSM (Higgs) \\\hline\hline
$\frac{Br(\mu^-\to e^-e^+e^-)}{Br(\mu\to e\gamma)}$  &  0.4\dots2.5  & $\sim6\cdot10^{-3}$ &$\sim6\cdot10^{-3}$  \\
$\frac{Br(\tau^-\to e^-e^+e^-)}{Br(\tau\to e\gamma)}$   & 0.4\dots2.3     &$\sim1\cdot10^{-2}$ & ${\sim1\cdot10^{-2}}$\\
$\frac{Br(\tau^-\to \mu^-\mu^+\mu^-)}{Br(\tau\to \mu\gamma)}$  &0.4\dots2.3     &$\sim2\cdot10^{-3}$ & $0.06\dots0.1$ \\
$\frac{Br(\tau^-\to e^-\mu^+\mu^-)}{Br(\tau\to e\gamma)}$  & 0.3\dots1.6     &$\sim2\cdot10^{-3}$ & $0.02\dots0.04$ \\
$\frac{Br(\tau^-\to \mu^-e^+e^-)}{Br(\tau\to \mu\gamma)}$  & 0.3\dots1.6    &$\sim1\cdot10^{-2}$ & ${\sim1\cdot10^{-2}}$\\
$\frac{Br(\tau^-\to e^-e^+e^-)}{Br(\tau^-\to e^-\mu^+\mu^-)}$     & 1.3\dots1.7   &$\sim5$ & 0.3\dots0.5\\
$\frac{Br(\tau^-\to \mu^-\mu^+\mu^-)}{Br(\tau^-\to \mu^-e^+e^-)}$   & 1.2\dots1.6    &$\sim0.2$ & 5\dots10 \\
$\frac{R(\mu\text{Ti}\to e\text{Ti})}{Br(\mu\to e\gamma)}$  & $10^{-2}\dots 10^2$     & $\sim 5\cdot 10^{-3}$ & $0.08\dots0.15$ \\\hline
\end{tabular}
\end{center}
\renewcommand{\arraystretch}{1.0}
}
\caption{\it Comparison between the LHT model and the MSSM without and with significant Higgs contributions.\label{tab:ratios}}
\label{tab:comparison}
\end{table}

The main messages from~\cite{BBDPT} are as follows:

\begin{itemize}
\item{Several rates can reach or approach the present
    experimental upper bounds. In particular, in order to suppress the $\mu \rightarrow e \gamma$ and $\mu^-
\rightarrow e^- e^+ e^-$ decay rates below the experimental upper bounds (see Fig.~\ref{fig:mue}), the
$V_{H\ell}$ mixing matrix has to be rather hierarchical, unless mirror
leptons are quasi-degenerate.} 
\item{The pattern of the LFV branching ratios in the LHT model differs
    significantly from the MSSM one, allowing a clear distinction of these two
    models.
The origin of this difference is that in the MSSM the LFV rates are dominated
    by the dipole operator, whose role is instead negligible in the LHT
    model.}
\item{These different patterns of LFV in the LHT and the MSSM can best be seen
    by studying certain correlations between branching ratios that have been
    previously considered in the context of the MSSM~\cite{Ellis:2002fe,Brignole:2004ah,Arganda:2005ji,Paradisi1}. 
We find that the ratios in Table~\ref{tab:comparison} could allow for a 
transparent distinction between the LHT model and the MSSM.
In particular, the ratios involving $Br(\ell_i \to \ell_j \gamma)$ turn out to
    be of $\mathcal{O}(1)$ and $\mathcal{O}(\alpha)$ in the LHT model and the
    MSSM, respectively.}
\item{We also note that a measurement of $\mu \to e \gamma$ at the $10^{-13}$
    level would necessarily imply within the MSSM a rate for the $\mu - e$
    conversion in Ti below $10^{-15}$, while it could be
    much larger within the LHT model.}
\item{Finally, we have studied the muon anomalous magnetic moment $(g-2)_\mu$,
and found that LHT effects are roughly a factor $5$ below the current
experimental uncertainty~\cite{Bennett:2006fi}, implying that the possible discrepancy between the SM prediction
and the data cannot be solved in the LHT model.
This represents another clear difference from the MSSM.} 
\end{itemize}

\section{Conclusions}
Our trilogy on FCNC processes in the LHT model revealed very interesting and
peculiar patterns that not only differ from those found in the SM and MFV
models but also in the MSSM.
These differences can most clearly be seen in LFV processes.
We are looking forward to the forthcoming data from Tevatron, LHC, $B-$, $K-$
and LFV dedicated experiments that will tell us whether the LHT model
represents a good description of nature.

\vspace*{-0.2cm}
\begin{acknowledgments}
\vspace*{-0.2cm}
We would  like to thank the {\it CKM2006} organizers for the wonderful hospitality
in Nagoya.
Special thanks go to the other authors of the trilogy:
Monika Blanke, Bj\"orn Duling, Anton Poschenrieder, Stefan
Recksiegel, Selma Uhlig, Andreas Weiler.
Partially Supported by BMBF under contract 05HT6WOA and by the Cluster of Excellence: Origin and Structure of the Universe. 
\end{acknowledgments}

\end{document}